Application of mixed-variable physics-informed neural networks to solve normalised momentum and energy transport equations for 2D internal convective flow

Ryno Laubscher [a, b], Pieter Rousseau [a]

[a] Department of Mechanical Engineering, Applied Thermal-Fluid Process Modelling Research Unit, University of Cape Town, Library Rd, Rondebosch, Cape Town, 7701, South-Africa

[b] Corresponding author E-mail address: ryno.laubscher@uct.ac.za

# Abstract

The prohibitive cost and low fidelity of experimental data in industry-scale thermofluid systems limit the usefulness of pure data-driven machine learning methods. Physics-informed neural networks (PINN) strive to overcome this by embedding the physics equations in the construction of the neural network loss function. In the present paper, the mixed-variable PINN methodology is applied to develop steady-state and transient surrogate models of incompressible laminar flow with heat transfer through a 2D internal domain with obstructions. Automatic spatial and temporal differentiation is applied to the partial differential equations for mass, momentum and energy conservation, and the residuals are included in the loss function, together with the boundary and initial values. Good agreement is obtained between the PINN and CFD results for both the steady-state and transient cases, but normalization of the PDEs proves to be crucial. Although this proves the ability of the PINN approach to solve multiple physics-based PDEs on a single domain, the PINN takes significantly longer to solve than the traditional finite volume numerical methods utilized in commercial CFD software.

Keywords: deep learning; physics-informed neural networks, computational fluid dynamics.

## 1. Introduction

The recent advances and achievements of machine learning are well-known and have significantly influenced a wide range of scientific fields such as agricultural sciences [1], neuroscience [2] and geosciences [3]. The application of machine learning methods to thermofluid engineering problems has also seen a recent surge in interest. Machine learning methods in engineering are typically employed to develop computationally efficient regression models with the capability to accurately fit highly non-linear, high-dimensional domains. The data used to train these models could either be experimental measurements or simulation results. The prohibitive cost and low fidelity of experimental data and high computational cost of industrial-scale simulation models are major challenges for current state-of-the-art machine learning methods [4] when the aim is to develop accurate predictive models (surrogate models) for thermofluid processes.

One approach to overcome this low data size problem is the implementation of prior knowledge into the learning process, such as the physical or empirical equations describing the underlying physics. Examples of such prior knowledge are the mass, momentum and energy conservation equations. The relatively new area of machine learning called physics-informed neural networks (PINN) [5] is one such approach which incorporates prior knowledge in the learning process by embedding the physics equations in the construction of the neural network loss function. Specific applications of PINNs include, but are not limited to, data-driven forecasting of physical processes, multi-scale modelling and control applications [4].

The present work applies the PINN methodology to develop steady-state and transient surrogate models of laminar flow through an internal domain. Therefore, the neural networks will be tasked to learn non-linear functions that accurately approximate the mass, momentum and energy conservation equations throughout the domain. The goal of the current paper is to investigate the accuracy of the PINN approach to learn fluid flow governing partial differential equations (PDEs) with respect to conventional finite volume method (FVM) results on a practical computational domain.

Several authors have applied machine learning techniques to thermofluid engineering problems. Typically, experimental data is used to develop the predictive models. For example: time-series forecasting of metal temperatures in a solid-fuel power plant [6], dynamic prediction of NOx and CO emissions [7], combustion optimization of coal-fired boiler [8], solar irradiation prediction [9] and prediction of critical heat fluxes in two-phase systems [10]. The disadvantage of using experimental data is that the complexity of the predictive model output is a function of the number of measurement locations, which is typically quite sparse in industrial thermofluid processes.

To develop machine learning models capable of predicting high-dimensional outputs, such as a stress distribution in a 3D solid body, simulation data is used to train the models. These predictive models trained using simulation data are also called surrogate models and are applied in various areas such as design space optimization, performance monitoring or real-time visualization. There have been numerous applications of machine learning in the development of thermofluid process surrogate models using simulation data. For example, the use of generative deep learning to predict species, temperature and velocity fields in a 2D flame [11], use of generative adversarial networks to recreate 2D pressure fields around aerofoils [12], to predict wake velocity and turbulence using multilayer perceptron networks [13] and prediction of thermal comfort in a vehicle using 3D CFD (computational fluid dynamics) data and neural networks [14]. Maulik et al. [15] developed a turbulent-eddy viscosity surrogate model using CFD training data. They employ the surrogate to bypass the solution of any extra PDEs for closure of the Reynolds-Averaged Navier-Stokes (RANS) simulations, thereby obtaining computational speed-up.

Although these data-driven surrogate models are quite capable of accurately mapping high-dimensional inputs to outputs, they require large amounts of data points. Acquiring these data points is often computationally expensive and requires careful design of experiments [16]. PINNs sets out to circumvent this shortcoming by constraining the neural network parameters (weights and biases) within a feasible solution space by embedding the conservation equations in the loss function.

Raissi et al. [4] introduced the concept of PINNs and showcased its ability to not only be used as predictive models, but also to discover physics. The authors used PINNs to develop accurate transient surrogate models of complex physics equations such as Schrodinger and Allen-Cahn. Zhu et al. [17] demonstrated physics-constrained surrogate modeling for stochastic PDEs trained without simulation data. The training is done by learning to solve the PDEs based on a loss function that contains both the residuals and the given boundary conditions. They included the conditional density of the predicted solution by distilling it from a Boltzmann-Gibbs reference density distribution. The model was applied to steady-state Darcy flow in random heterogeneous media. They obtained high predictive accuracy for high-dimensional stochastic input fields and found that the generalization performance of the proposed physics-constrained surrogate is consistently better than data-driven surrogates for out-of-distribution test inputs.

Mao et al. [18] applied PINNs to approximate the Euler equation to model high-speed aerodynamic flows in 1D and 2D. Rao et al. [16] developed a PINN surrogate model capable of accurately predicting laminar flow around a cylinder. They also proposed a mixed-variable PINN approach to learn the momentum conservation equation by separately predicting the Cauchy stress tensor, which significantly enhanced the predictive accuracy of the network. The mixed-variable PINN approach does not require the use of automatic differentiation to calculate second-order derivatives, which is the main driver for the increase in predictive accuracy. Jagtap et al. [19] investigated the use of adaptive activation functions to speed-up the training process of PINNs. The authors found a significant speed-up in training a surrogate model of the transient Schrodinger equation on a 1D domain. Sun et al. [20] studied the accuracy and sensitivity of PINNs when predicting flows in simple converging-diverging ducts.

Meng et al.[21] developed a parareal or parallel-in-time PINN (PPINN) for transient PDEs. A coarse-grained solved is used to solve a simplified PDE for the entire time-domain to obtain an initial solution. Fine-grained PINNs are then employed in parallel for a series of time sub-domains based on the initial conditions. This is done in an iterative process and avoids the prohibitive training cost associated with very large time-space computational domains when long time integration is sought. They applied the PPINN to the one-dimensional Burgers equation for viscous flow as well as to a two-dimensional diffusion reaction equation, both under isothermal conditions.

In the present paper, the mixed-variable PINN methodology is applied to develop steady-state and transient surrogate models of incompressible laminar flow with heat transfer through a 2D internal domain with heated cylindrical obstructions (tubes). The paper investigates the effects of hyperparameters such as the number of hidden layers, neurons per hidden layer and number of coordinate points on model accuracy. *Python 3.6.5* and *Tensorflow 1.15.0* are used in the current work to develop the PINN model and to process the input data and results.

## 2. Materials and methods

### 2.1 Physics equations

In the present work, the PINN scheme is applied to solve the physics-based partial differential equations (PDEs) for mass, momentum and energy conservation in incompressible laminar flow. We consider these PDEs assuming constant fluid properties (density, viscosity, specific heat and thermal conductivity) [22].

Conservation of mass (continuity):

$$\nabla \cdot \bar{u} = 0 \tag{1}$$

Conservation of momentum (Navier-Stokes):

$$\rho \frac{\partial \bar{u}}{\partial t} + \rho (\bar{u} \cdot \nabla) \bar{u} = -\nabla p + \mu \nabla \cdot (\nabla \bar{u}) \tag{2}$$

Conservation of energy:

$$\rho c_p \frac{\partial T}{\partial t} + \rho \nabla \cdot (\bar{u} c_p T) = \lambda \nabla \cdot (\nabla T) \tag{3}$$

To apply the mixed-variable PINN approach proposed by Rao et al. [16], the above equations are written in a continuum and constitutive form that eliminates the need to directly calculate higher-order derivatives. It should be noted the second-order derivatives are still taken of the stream function to calculate the velocity gradients as will be shown lateron in the paper. This has been shown to improve the trainability of the neural networks. The neural networks therefore not only predict the latent solution variables such as temperature, velocity and pressure, but also the stress tensor components and directional diffusive heat fluxes.

The 2D momentum equations in its derivative forms are given by

$$\rho \frac{\partial u}{\partial t} + \rho \left( u \frac{\partial u}{\partial x} + v \frac{\partial u}{\partial y} \right) = \frac{\partial \sigma_{11}}{\partial x} + \frac{\partial \sigma_{12}}{\partial y}$$
$$\rho \frac{\partial v}{\partial t} + \rho \left( u \frac{\partial v}{\partial x} + v \frac{\partial v}{\partial y} \right) = \frac{\partial \sigma_{12}}{\partial x} + \frac{\partial \sigma_{22}}{\partial y}$$
(4)

The Cauchy stress tensor components are given by

$$\sigma_{11} = -p + 2\mu \frac{\partial u}{\partial x}$$
$$\sigma_{22} = -p + 2\mu \frac{\partial v}{\partial y}$$
$$\sigma_{12} = \mu \left( \frac{\partial u}{\partial y} + \frac{\partial v}{\partial x} \right)$$
$$p = -\frac{\sigma_{11} + \sigma_{22}}{2}$$
(5)

The higher-order derivatives of the 2D energy conservation equation can similarly be removed. The energy conservation is given by

$$\rho c_P \frac{\partial T}{\partial t} + \rho c_P \left( u \frac{\partial T}{\partial x} + v \frac{\partial T}{\partial y} \right) = \frac{\partial q_x}{\partial x} + \frac{\partial q_y}{\partial y}$$
(6)

The diffusive heat fluxes in the x- and y-directions are given by

$$q_x = \lambda \frac{\partial T}{\partial x}$$
$$q_y = \lambda \frac{\partial T}{\partial y}$$
(7)

The following sections provide a short overview of neural networks and PINNs followed by the detail model development, which covers aspects such as pre-processing of the PDEs, neural network loss calculation, boundary conditions and the computational domain under consideration.

### 2.2 Multilayer perceptron neural networks

Multilayer perceptron (MLP) networks are the archetypical form of artificial neural networks [23]. These are mainly applied in supervised learning applications where input variables are accurately mapped to target variables by optimizing the weights $\overline{w}$ and biases $\overline{b}$. An MLP network consists of three sections namely the input layer, hidden layers and output layer. Each layer of the neural network consists of multiple computing units called neurons. The output signal $s$ from a neuron is fed to each neuron in the downstream layer of the network, where it is multiplied by a weight and summed with other weighted signals. The summed weighted signal is then passed into a nonlinear activation function which generates the layer output signal vector. This process is repeated until the signal reaches the output layer. The operation for a single layer $l$ is mathematically expressed by equation (8), where the layer has $n_l$ neurons.

$$\bar{s}_l = \bar{h}_{l-1} \cdot \overline{W}_l + \bar{b}_l \tag{8}$$

In equation (8), $\bar{h}_{l-1}$ is the output vector from the previous layer $l-1$, which has a vector length equal to $n_{l-1}$ (number of neurons in the previous layer), $\overline{W}_l$ is the matrix containing all the layer connection weights of size $n_{l-1} \times n_l$ and $\bar{b}_l$ is the bias vector of layer $l$ with length $n_l$.

A wide range of activation functions are available, but in the present work only linear and hyperbolic-tangent [24] functions will be utilized as shown in equations (9) and (10) respectively.

$$\bar{h}_l = \sigma_{linear}(\bar{s}_l) = \bar{s}_l = \bar{h}_{l-1} \cdot \overline{w}_l + \bar{b}_l \tag{9}$$

$$\bar{h}_l = \sigma_{tanh}(\bar{s}_l) = \frac{\exp(\bar{s}_l) - \exp(-\bar{s}_l)}{\exp(\bar{s}_l) + \exp(-\bar{s}_l)} \tag{10}$$

The output signal $\bar{h}_L$ from the final layer $L$ is the predicted target value $\bar{Z}$ of the MLP network. To ensure that the difference between the predicted and the actual target values is small, the trainable parameters of each layer, namely $\overline{W}$ and $\bar{b}$ must be optimized by minimization of a selected loss function $J(\bar{Z})$. For typical supervised learning applications, the mean squared error (MSE) between the actual and predicted values is used. For PINNs the loss function is configured differently and will be described in section 2.3. To minimize the loss function, gradient-based optimization is applied which requires the gradients of the loss function with respect to the trainable parameters $\nabla_{\overline{W},\bar{b}} J$ to be calculated. These gradients are determined using automatic differentiation, also called the back-propagation algorithm [25]. Once the gradients are known, the trainable parameters can be iteratively updated to minimize the loss function. In the present work, a first-order algorithm is implemented to train the network for the first 10000 iterations and then a quasi-Newton algorithm is used to train the network till convergence. The first-order optimization routine utilized is the well-known Adam algorithm [26] and the quasi-Newton optimization routine used is the Limited-memory Broyden-Fletcher-Goldfarb-Shanno (L-BFGS) method [27].

### 2.3 Physics-informed neural networks

In a PINN scheme, a deep MLP network is utilized as a transfer function to map spatiotemporal fields $[x, y, t]$ to latent solution variables $\phi$ representing physical fields such as pressure and temperature. In order to achieve this, a residual function $f$ is constructed from the physics-based PDEs of the form:

$$f(x,y,t) \approx \frac{\partial \phi(x,y,t)}{\partial t} + \Pi[\phi(x,y,t)] = 0 \tag{11}$$

In equation (11), $\Pi[\cdot]$ is the nonlinear differential operator which can represent a wide range of physics equation quantities. In the present work, the operator is used to represent the advection and diffusion quantities of the momentum and energy conservation equations. The different residual functions are then collected to form a combined residual loss functions ($J_{res}$).

To ensure that the solution of the MLP approximates the desired physical conditions, additional loss functions are added [28] to impose boundary conditions ($J_{bc}$) and initial conditions ($J_{init}$). The combined loss funcions are then added together to obtain and overall loss function as follows:

$$J_{loss} = J_{res} + \beta J_{bc} + J_{init} \tag{12}$$

In equation (12) $\beta$ is a user-defined weighting coefficient for the boundary condition losses and is typically set to a value of 2 [16]. The combined residual loss $J_{res}$, boundary condition loss $J_{bc}$ and the initial condition loss $J_{init}$ are given by

$$\begin{aligned} J_{res} &= \frac{1}{N_f} \sum_{i=1}^{N_f} \left[ f(x_i, y_i, t_i) \right]^2 \\ J_{bc} &= \frac{1}{N_{bc}} \sum_{j=1}^{N_{bc}} \left[ \phi(x_j, y_j) - \phi^{bc}_j \right]^2 \\ J_{init} &= \frac{1}{N_{init}} \sum_{q=1}^{N_{init}} \left[ \phi(x_q, y_q, 0) - \phi^{init}_q \right]^2 \end{aligned} \tag{13}$$

The minimization of $J_{res}$ during optimization of the neural network trainable parameters (weights and biases) will ensure that the solution satisfies the desired physics-based differential equations. To determine the quantities required to construct $f$, such as the temporal derivative $\partial \phi(x,y,t)/\partial t$ and the nonlinear derivatives $\Pi[\phi(x,y,t)]$, automatic differentiation [27] and the graph structure of the PINN are utilized. In equation (13), the $i$ subscript indicates spatiotemporal coordinates that lie within the computational domain.

$J_{bc}$ is minimized to ensure the calculated solution variables $\phi$ positioned on the boundaries of the computational domain are equal to the user-defined boundary condition values $\phi^{bc}$. Similarly, the initial condition loss $J_{init}$ is minimized to ensure the calculated solution variables are equal to the user-specified initial conditions $\phi^{init}$ at time $t = 0$. The $j$ subscript indicates the spatial coordinates that lie on the boundaries (inlets, outlets and walls) of the domain and the $q$ subscript indicate spatial locations used to evaluate the initial conditions.

For the PINN being developed in the current work there will be two residual loss terms, namely for the momentum $J_{mom,res}$ and energy $J_{T,res}$ conservation equations. The implemented form of the loss functions for the present work will be discussed in the model development section (2.4).

### 2.4 Problem setup

#### 2.4.1 Domain and boundary conditions

The proposed PINN methodology is applied to a rectangular internal flow domain with two cylindrical obstructions, as shown in figure 1. The duct and cylinder walls are heated, with a cooler fluid entering the domain through the inlet boundary. The inlet X and Y directional velocity components are specified to be $u_{inlet} = 4u_{max}(H-y)y/H^2$, $u_{max} = 1.0$ m/s, $v_{inlet} = 0$ m/s respectively, the outlet gauge pressure is set to $p_{outlet} = 0$ Pa, the inlet fluid temperature is set to $T_{inlet} = 300$ K and the temperature of the walls is set to

$T_{walls} = 600$ K. For the transient simulation the initial velocity and pressure is set to $u_{init} = v_{init} = 0$ m/s, $p_{init} = 0$ Pa and the initial temperature of the entire domain equal to the inlet fluid temperature, $T_{init} = T_{inlet}$. To simplify the current PINN implementation all fluid properties are assumed to be constant with: $\mu = 0.02$ kg/m·s, $c_p = 1000$ J/kg·K, $\lambda = 0.03$ W/m·K and $\rho = 1$ kg/m³.

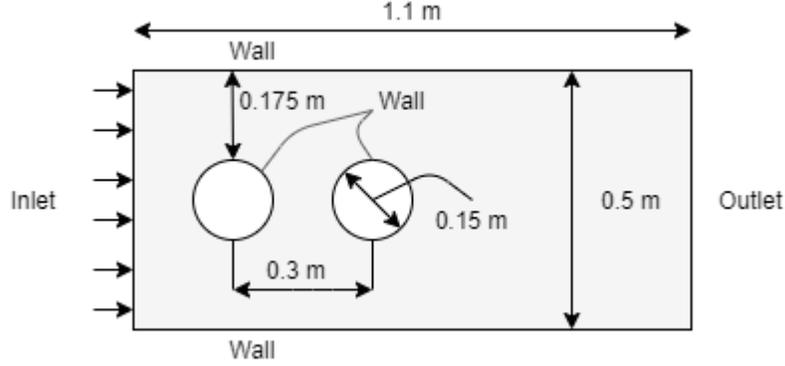

Figure 1: Sketch of the computational domain

### 2.4.2 Model development

A large difference in magnitude between the $J_{T,res}$ and $J_{mom,res}$ loss function components would result in the optimizer not effectively minimizing all the imposed physics equations. For example, in the current problem there is a significant difference in the magnitude of the transported scalars of the momentum ($u, v \sim 0 \rightarrow 1$ m/s) and energy equations ($T \sim 300 \rightarrow 600$ K). This effectively biases the loss function minimization towards the energy conservation equation residuals and boundary conditions while neglecting the momentum conservation equation residuals and boundary conditions. This, in turn, leads to extended training times and inaccurate results. To circumvent this, the conservation equations are normalised before implementation in the PINN.

The normalised (non-dimensional) input and solution variables are defined as:

$$x^* = \frac{x}{L}, \quad y^* = \frac{y}{L}, \quad u^* = \frac{u}{u_\infty}, \quad v^* = \frac{v}{u_\infty}, \quad p^* = \frac{p}{\rho u_\infty^2}, \quad T^* = \frac{T}{T_\infty}, \quad t^* = \frac{t}{\tau} \qquad (14)$$

In equation (14) the superscript $*$ denotes the non-dimensional variables. The spatial coordinates $x, y$ are normalized via the characteristic length $L$ of the domain, which was selected as $1.1$ m. The x and y velocity components $u, v$ are normalized using the maximum velocity at the inlet boundary $u_\infty = u_{max}$, the pressures $p$ are normalized via the maximum dynamic pressure $\rho u_\infty^2$, the temperatures $T$ are normalized via the fixed wall temperature $T_\infty = T_{walls}$, and the time $t$ is normalized via the minimum residence time within the domain, namely $\tau = \frac{L}{u_\infty}$.

From equation (4) we obtain the non-dimensional momentum conservation equations as

$$\frac{\partial u^*}{\partial t^*} + u^* \frac{\partial u^*}{\partial x^*} + v^* \frac{\partial u^*}{\partial y^*} = \frac{\partial \sigma_{11}^*}{\partial x^*} + \frac{\partial \sigma_{12}^*}{\partial y^*}$$
$$\frac{\partial v^*}{\partial t^*} + u^* \frac{\partial v^*}{\partial x^*} + v^* \frac{\partial v^*}{\partial y^*} = \frac{\partial \sigma_{12}^*}{\partial x^*} + \frac{\partial \sigma_{22}^*}{\partial y^*}$$

(15)

with

$$\sigma_{11}^* = \frac{\sigma_{11}}{\rho u_\infty^2} = -p^* + \frac{2}{\text{Re}} \frac{\partial u^*}{\partial x^*}$$

$$\sigma_{22}^* = \frac{\sigma_{22}}{\rho u_\infty^2} = -p^* + \frac{2}{\text{Re}} \frac{\partial v^*}{\partial y^*}$$

$$\sigma_{12}^* = \frac{\sigma_{12}}{\rho u_\infty^2} = \frac{1}{\text{Re}} \left( \frac{\partial u^*}{\partial y^*} + \frac{\partial v^*}{\partial x^*} \right)$$

$$p^* = -\frac{\sigma_{11}^* + \sigma_{22}^*}{2}$$

(16)

From equation (6) we obtain the non-dimensional energy conservation equation as

$$\frac{\partial T^*}{\partial t^*} + u^* \frac{\partial T^*}{\partial x^*} + v^* \frac{\partial T^*}{\partial y^*} = \frac{1}{\text{Re}\,\text{Pr}} \left( \frac{\partial q_x^*}{\partial x^*} + \frac{\partial q_y^*}{\partial y^*} \right)$$

(17)

with

$$q_x^* = \frac{q_x}{\lambda T_\infty / L} = \frac{\partial T^*}{\partial x^*}$$

$$q_y^* = \frac{q_y}{\lambda T_\infty / L} = \frac{\partial T^*}{\partial y^*}$$

(18)

In equations (16) and (17) the non-dimensional Reynolds number is defined as $\text{Re} = \rho u_\infty L / \mu$ and the non-dimensional Prandtl number as $\text{Pr} = c_p \mu / \lambda$.

Mass conservation is enforced by solving for the non-dimensional stream function $\psi$ while satisfying the definition shown in equation (19). This ensures a divergence-free flow field, which automatically satisfies the mass conservation equation (1) for incompressible flow.

$$u^* = \frac{\partial \psi}{\partial y^*}, \quad v^* = -\frac{\partial \psi}{\partial x^*}$$

(19)

The magnitudes of the non-dimensional velocity components are calculated from the stream function definition via automatic differentiation. This eliminates the need for an additional residual loss function to be minimized by the PINN during training.

As shown in figure 2, the neural network uses the non-dimensional spatial and temporal parameters as inputs variables and provides eight non-dimensional solution variables.

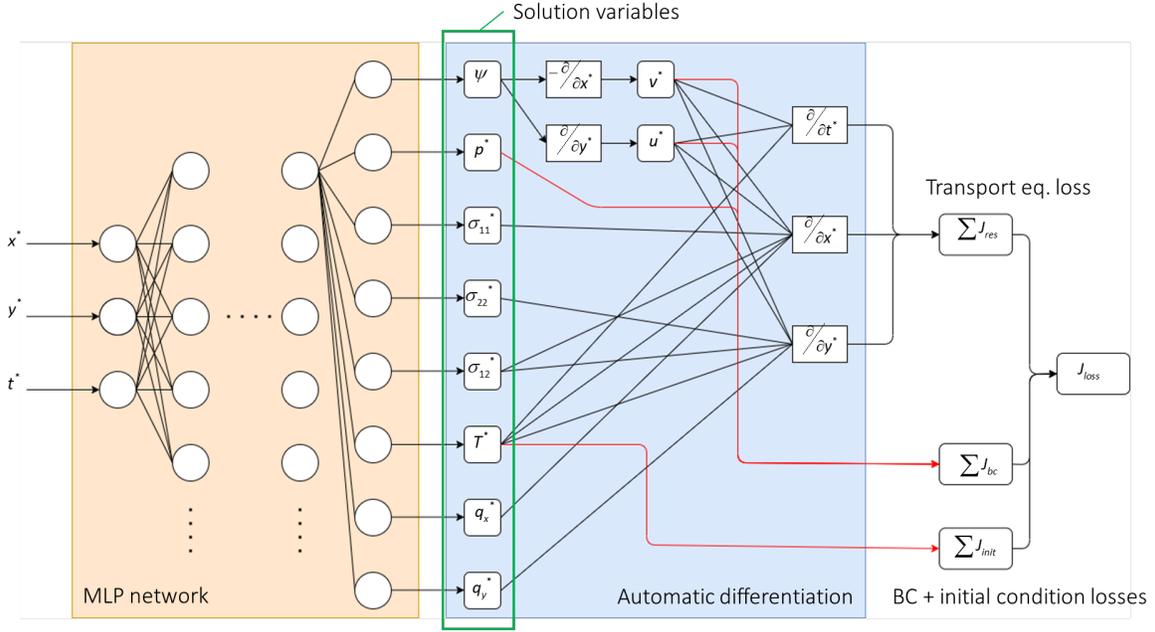

Figure 2: Architecture of PINN developed in current work. The red lines indicate the flow of solution variables directly to the boundary and initial conditions loss calculations. The calculated loss function $\sum J_{loss}$ is minimized using Adam and L-BFGS optimization methods. $\sum J_{res}$ is the combined residual function loss, $\sum J_{bc}$ is the combined boundary condition loss and $\sum J_{init}$ is the combined initial condition loss.

The outputs from the MLP network, shown in figure 2, are fed to the automatic differentiation block which calculates the required spatial and temporal gradients to approximate the various residual functions $f$ [equation (11)]. The residual functions for the x- and y-momentum conservation and the Cauchy stress tensor components are calculated using equations (20) and (21).

$$f_{x-mom} = \frac{\partial u^*}{\partial t^*} + u^* \frac{\partial u^*}{\partial x^*} + v^* \frac{\partial u^*}{\partial y^*} - \left[ \frac{\partial \sigma_{11}^*}{\partial x^*} + \frac{\partial \sigma_{12}^*}{\partial y^*} \right]$$

$$f_{y-mom} = \frac{\partial v^*}{\partial t^*} + u^* \frac{\partial v^*}{\partial x^*} + v^* \frac{\partial v^*}{\partial y^*} - \left[ \frac{\partial \sigma_{12}^*}{\partial x^*} + \frac{\partial \sigma_{22}^*}{\partial y^*} \right]$$

(20)

$$f_{\sigma 11} = -p^* + \frac{2}{Re} \frac{\partial u^*}{\partial x^*} - \sigma_{11}^*$$

$$f_{\sigma 22} = -p^* + \frac{2}{Re} \frac{\partial v^*}{\partial y^*} - \sigma_{22}^*$$

$$f_{\sigma 12} = \frac{1}{Re} \left( \frac{\partial u^*}{\partial y^*} + \frac{\partial v^*}{\partial x^*} \right) - \sigma_{12}^*$$

$$f_p = p^* + \frac{\sigma_{11}^* + \sigma_{22}^*}{2}$$

(21)

The residual functions for the energy conservation equation and the conductive heat fluxes are calculated using equations (22) and (23).

$$f_T = \frac{\partial T^*}{\partial t^*} + u^* \frac{\partial T^*}{\partial x^*} + v^* \frac{\partial T^*}{\partial y^*} - \frac{1}{\text{RePr}} \left( \frac{\partial q_x^*}{\partial x^*} + \frac{\partial q_y^*}{\partial y^*} \right) \tag{22}$$

$$f_{qx} = q_x^* - \frac{\partial T^*}{\partial x^*}$$
$$f_{qy} = q_y^* - \frac{\partial T^*}{\partial y^*} \tag{23}$$

The combined momentum residual loss function $J_{mom,res}$ is calculated using equation (24).

$$J_{mom,res} = \frac{1}{N_f} \sum_{i=1}^{N_f} \left[ f_{x-mom}^2 + f_{y-mom}^2 + f_{\sigma 11}^2 + f_{\sigma 22}^2 + f_{\sigma 12}^2 + f_p^2 \right] \tag{24}$$

Similarly, the combined energy residual loss function $J_{T,res}$ is calculated using equation (25).

$$J_{T,res} = \frac{1}{N_f} \sum_{i=1}^{N_f} \left[ f_T^2 + f_{qx}^2 + f_{qy}^2 \right] \tag{25}$$

Minimizing the summation of equations (24) and (25) ($\sum J_{res} = J_{mom,res} + J_{T,res}$, see figure 2) during neural network training, will ensure that the learnt transfer function approximates the underlying physics-based conservation equations. The conservation equation residual loss functions are only evaluated at physical locations that lie within the internal flow domain.

The boundary condition loss functions ensure that the calculated solution variables on the boundaries are equal to the user-defined values. Assuming a no-slip condition at the walls and implementing the abovementioned inlet and outlet boundary conditions, the momentum boundary condition loss function can be calculated using equation (26).

$$J_{mom,bc} = \frac{1}{N_{bc,inlet}} \sum_{j=1}^{N_{bc,inlet}} \left[ u^*(x_j, y_j) - \frac{u_{inlet}}{u_\infty} \right]^2 + \frac{1}{N_{bc,inlet}} \sum_{j=1}^{N_{bc,inlet}} \left[ v^*(x_j, y_j) - \frac{v_{inlet}}{u_\infty} \right]^2 + \\ \frac{1}{N_{bc,outlet}} \sum_{j=1}^{N_{bc,outlet}} \left[ p^*(x_j, y_j) - \frac{p_{outlet}}{\rho u_\infty^2} \right]^2 + \frac{1}{N_{bc,walls}} \sum_{j=1}^{N_{bc,walls}} \left[ u^*(x_j, y_j) \right]^2 + \frac{1}{N_{bc,walls}} \sum_{j=1}^{N_{bc,walls}} \left[ v^*(x_j, y_j) \right]^2 \tag{26}$$

The temperature boundary conditions are enforced by minimizing the loss function in equation (27).

$$J_{T,bc} = \frac{1}{N_{bc,inlet}} \sum_{j=1}^{N_{bc,inlet}} \left[ T^*(x_j, y_j) - \frac{T_{inlet}}{T_{walls}} \right]^2 + \frac{1}{N_{bc,walls}} \sum_{j=1}^{N_{bc,walls}} \left[ T^*(x_j, y_j) - \frac{T_{walls}}{T_{walls}} \right]^2 \tag{27}$$

The combined boundary condition loss function is then calculated as $\sum J_{bc} = J_{T,bc} + J_{mom,bc}$ (see figure 2). In equations (26) and (27), it is assumed that the boundary conditions do not vary over time. The boundary condition loss functions are only evaluated at locations lying on the domain boundaries.

For transient simulations, the initial conditions are imposed on the pressure, velocity and temperature solution variables. The initial condition loss function for the momentum variables is calculated using equation (28).

$$J_{mom,init} = \frac{1}{N_{init}} \sum_{q=1}^{N_{init}} \left[ u^*(x_q, y_q, 0) - \frac{u_{init}}{u_\infty} \right]^2 + \frac{1}{N_{init}} \sum_{q=1}^{N_{init}} \left[ v^*(x_q, y_q, 0) - \frac{v_{init}}{u_\infty} \right]^2 + \frac{1}{N_{init}} \sum_{q=1}^{N_{init}} \left[ p^*(x_q, y_q, 0) - \frac{p_{init}}{\rho u_\infty^2} \right]^2 \quad (28)$$

Similarly, the initial condition loss function for the temperature solution is calculated using equation (29).

$$J_{T,init} = \frac{1}{N_{init}} \sum_{q=1}^{N_{init}} \left[ T^*(x_q, y_q, 0) - \frac{T_{init}}{T_{walls}} \right]^2 \quad (29)$$

The combined initial condition loss is then calculated as $\sum J_{init} = J_{T,init} + J_{mom,init}$.

The total loss of the PINN is used to optimize the trainable network parameters and is calculated as the sum of the combined conservation equation residual, boundary condition and initial condition losses: $\sum J_{loss} = \beta \left( \sum J_{bc} + \sum J_{init} \right) + \sum J_{res}$.

### 2.4.3 Geometrical framework for the PINN

The various loss functions described above are evaluated at different spatial and temporal coordinates in the computational domain. The conservation equation residual loss functions are only evaluated at physical locations lying within the internal flow domain, while the boundary condition loss functions are only evaluated at the inlet, outlet and wall boundary locations. We therefore need to consider the geometrical support framework for the implementation of the PINN, which does not rely on a conventional physical framework such as the computational mesh used in the CFD finite volume approach.

In the CFD approach the continuous domain is replaced by a discrete mesh with non-overlapping cells and faces. The partial derivatives are then approximated via discretization and transformed to an equivalent set of algebraic equations that must be satisfied for each of the cells in the computational domain. These algebraic equations for each cell also contain the values of the solution within the neighbouring cells. Therefore, the geometry and topology of the cells, i.e. its dimensions and where the cells are located relative to one another, are crucial inputs.

In the PINN the derivatives are evaluated exactly via automatic differentiation. This is achieved by applying the chain rule repeatedly to each of the elementary arithmetic operations and functions that form part of the trained MLP transfer function. Although the spatial coordinate positions of the selected points are required in this process, there is no concept of a cell volume and the location of the points relative to one another is not required. This fundamental difference between the two approaches may offer several benefits, the discussion of which is outside the scope of this paper.

In order to construct a representative grid of spatial coordinates for the PINN a fine-resolution numerical mesh of the computational domain was created using ANSYS R3® Design Modeler and Mesher. The mesh had 120000 internal cells, 440 inlet and outlet faces and 1200 wall faces. For the internal cells, the cell centroids were extracted and for the boundaries, the face centres were extracted and written to a file. The colocation coordinates for use in the PINN were then sampled from the coordinates file using the Latin hypercube sampling method. Figure 3 shows an example of sampled coordinate points in the PINN computational domain.

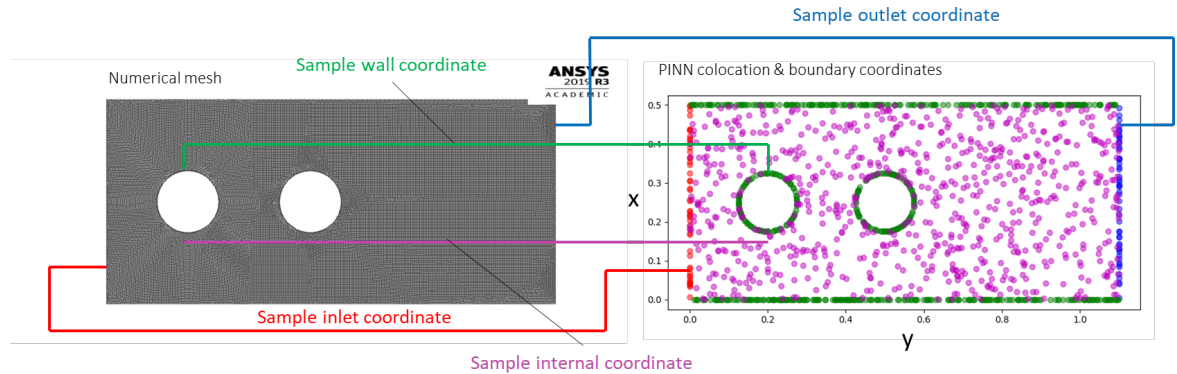

Figure 3: Colocation coordinates sampled from the numerical mesh. Red – inlet boundary coordinates, blue – outlet boundary coordinates, green – wall boundary coordinates, purple – internal colocation coordinates

## 3. Results and discussion

### 3.1 Hyperparameter search results

To find the best performing PINN configuration a coarse grid search was employed. The hyperparameter search was applied using the steady-state PINN model and investigated the effect of the number of colocation points, the number of hidden layers and the number of neurons per layer on the PINN total loss value $\sum J_{loss}$ . For the developed PINN models the neural network hidden layers used hyperbolic-tangent activation functions and the output layers used linear activation functions. The learning rate of the Adam algorithm was set to 0.001 and batch gradient descent was implemented.

To investigate the effect of the number of colocation points on the calculated PINN total loss value, a fixed neural network architecture of 6 hidden layers and 64 neurons per hidden layer was used. The number of colocation points used during PINN training was varied between 1500-96000. For each evaluated dataset size, the PINN model was trained and the total loss recorded, the results of which can be seen in figure 4.

The results in figure 4 (left) show that there is a drop (approximately 250%) in the total loss value when comparing the use of 1500 and 6000 colocation points. The loss value does not change significantly between 6000 and 96000 colocation points, and a minimum loss of 0.00488 is found when using 24000 colocation coordinates. The convergence curves with respect to the number of colocation points used during training are also shown in figure 4 (right). The results show that the rate of convergence of the different models is very similar except for the model trained using 24000 points, which converges slightly faster than the other models.

The conservation equations were also solved in their original dimensional form using the same PINN architecture with 24000 colocation coordinates. This was done to gauge the performance improvement of the non-dimensionalized approach. For the dimensional approach the model loss converged to a value of

9E5 which is orders of magnitude larger than the non-dimensional approach. The major contributor to the large loss value was the wall temperature boundary conditions (99.5% of total loss value).

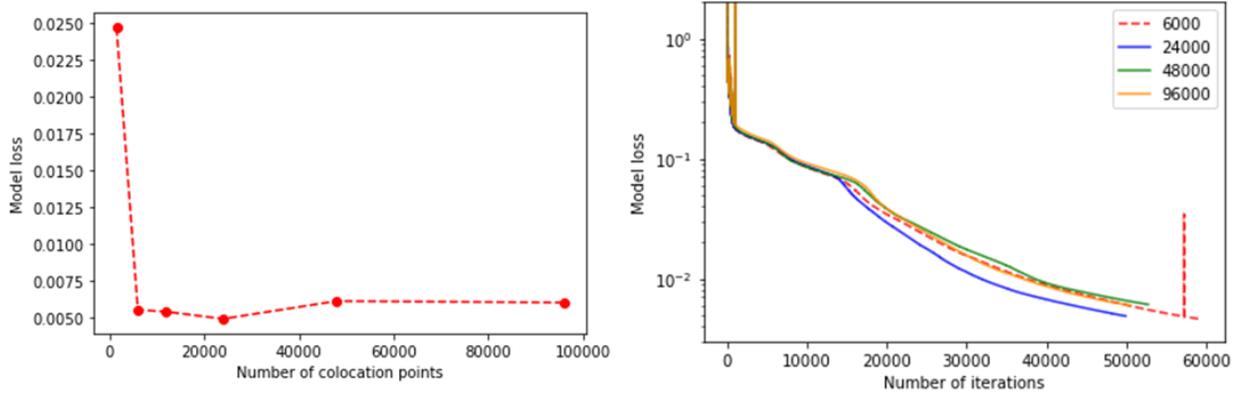

Figure 4: Effect of the number of colocation points on model loss and training history for the fixed architecture of 6x64 neural network.

The effect of number of neurons per hidden layer and number of hidden layers on the total loss were investigated while keeping the number of sampled colocation points constant at 24000. The number of hidden layers was 3, 6, and 9 and for each hidden layer depth configuration the number of neurons per layer was 32, 64 and 128. The results for the architecture hyperparameter grid search are shown in table 1.

Table 1: Model total losses for different architectures

| | Number of hidden layers | | |
|---|---|---|---|
| **Number of neurons per layer** | **3** | **6** | **9** |
| **32** | 0.0931 | 0.010707 | 0.004962 |
| **64** | 0.076423 | 0.004886 | 0.001472 |
| **128** | 0.0237 | 0.00311 | 0.000994 |

The results in table 1 show that increasing the number of neurons per hidden layer or the depth of the neural network reduces the resulting loss value. The depth of the neural network had a more pronounced effect than the number of neurons.

In order to validate the proposed PINN methodology the same problem was also analysed with the aid of ANSYS Fluent® R3 using the finite volume CFD approach applied to the fine-resolution numerical mesh described above. In the following sections the results of the PINN model with 9 layers and 128 neurons per hidden layer will be compared to the benchmark CFD results.

### 3.2 Steady-state model results

Figure 5 shows contour maps of the velocity magnitude, static pressure and temperature results of the steady-state mixed-variable PINN model along with the reference CFD model results. Figure 5 also shows three vertical line probes of these flow quantities at X coordinates of 0.1, 0.3 and 0.8 m. The results have been rescaled to their respective dimensional quantities using the relations in equation (14).

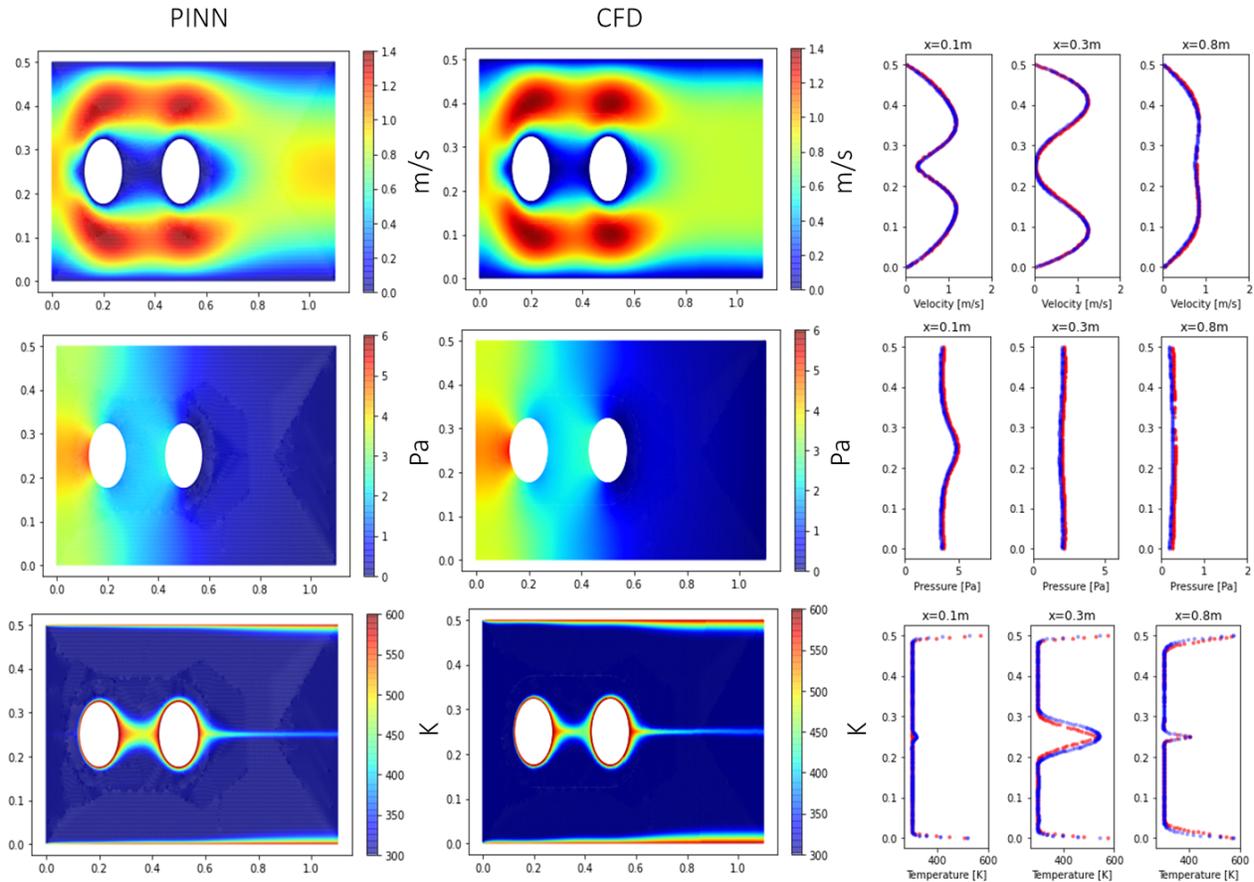

Figure 5: Contours and line probes of steady-state CFD and PINN results. Blue: PINN, Red: CFD

When compared to the CFD model results the PINN model captures the velocity distribution with reasonable accuracy. The predicted CFD velocity profile at the exit boundary away from the wall is nearly uniformly spread across the height of the domain, whereas the PINN profile has a slightly more parabolic shape. This effectively means that the velocity boundary layer is thicker in the case of the PINN than for the CFD. The static pressure in between the two cylinders predicted by the PINN model is slightly lower when compared to the reference solution. For the remainder of the domain, the PINN pressure distribution is in close agreement with the reference model.

The PINN model predicts a slightly more diffuse vertical temperature profile between the two cylinders when compared to the CFD results and a thinner thermal boundary layer on the upper and lower horizontal walls at x = 0.6-1.1 m. However, the PINN model captures the temperature distribution in the trailing stream of hot fluid that leaves the last cylinder surface with good accuracy. A possible cause for the prediction of a thicker velocity and thinner thermal boundary layer near the exit of the domain may be that the PINN model does not have enough colocation points in the near wall zone and exit plane to accurately resolve the profiles. To investigate this the 9-layer model was retrained using 96000 colocation points and the velocity and temperature contours plotted, as shown in figure 6. The results below show that the PINN model still predicts a thicker velocity boundary near the exit of the domain, but the thermal boundary layer on the horizontal walls are in closer agreement with the CFD results.

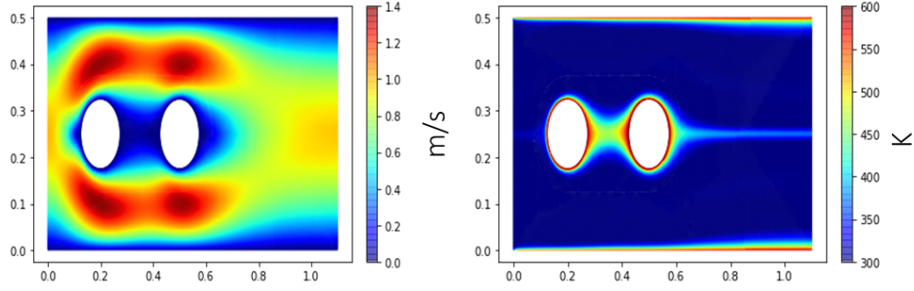

Figure 6: PINN velocity and temperature contour plots when training model with 96000 colocation points

It is interesting to note that for the specific problem the convergence criteria used in Fluent for the CFD benchmark had to be tightened to a continuity residual of 1E-5, velocity 2E-6 and energy 4E-8 in order to obtain sufficiently converged results, whereas the default values are typically 1E-3 for continuity, 1E-4 for velocity and 1E-6 for the energy equation.

The mean absolute percentage error (MAPE) between the CFD and PINN model results for the steady-state case was calculated using equation (30). In equation (30), $\phi$ is a placeholder for the solution variables namely velocity, static pressure and temperature and $N$ is the total number of colocation coordinates where the error is evaluated between the CFD and PINN models. The steady-state MAPEs can be seen in table 2.

$$MAPE = \frac{\sum_{i=0}^{N}\left|\phi_{PINN}^{i} - \phi_{CFD}^{i}\right|}{\left|\phi_{CFD}^{i}\right|} \cdot 100\% \tag{30}$$

Table 2: Mean absolute error of steady-state model results (PINN model trained with 96000 colocation points)

| Solution variable | Velocity magnitude | Static pressure | Temperature |
|---|---|---|---|
| MAPE [%] | 4.6 | 6.1 | 2.2 |

### 3.3 Transient model results

The 2D grid points shown in figure 3 that were used for the steady-state model are only for a single time step. For the transient analysis these points were not fixed, but rather similar 2D slices of x and y coordinates were sampled at specified time steps and stacked on top of each other to form a 3D dataset with the axes corresponding to the x, y and time dimensions. Figure 7 below visualizes these sampled coordinates. The figure on the left shows the various boundary conditions along with the initial conditions and the figure on the right the sampled internal coordinates (blue dots).

For the transient PINN model, approximately 110000 internal domain, 5000 inlet boundary, 6500 outlet boundary, 23000 wall boundary and 4000 initial condition colocation coordinates were sampled. In the time dimension, 80 steps were sampled over the 1.0 second period using Latin hypercube sampling. The initial conditions for the velocity vector components and static pressure were set to 0 m/s and 0 Pa respectively. For the temperature, the entire domain was set to 300 K. Similar to the work of Rao et al. [16], a time-variant inlet velocity profile was applied which is defined as shown in equation (31).

$$u_{inlet}(0,y,t) = 4u_{max}(H-y)y/H^2 \cdot \left( \sin\left[ \frac{2\pi t^*}{2\cdot\tau} + \frac{3\pi}{2} \right] + 1 \right) \tag{31}$$

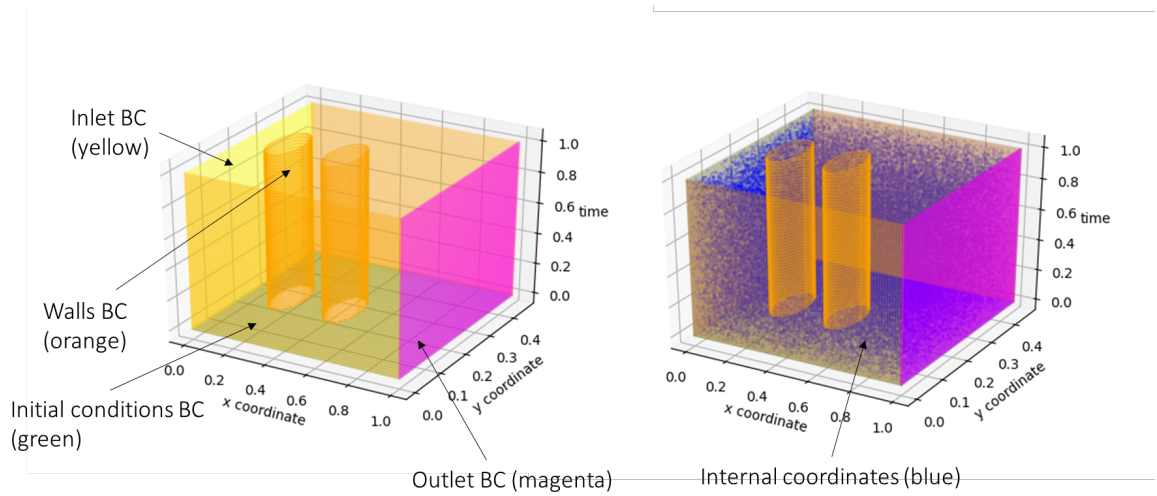

Figure 7: Visualization of sampled coordinates for transient dataset

Figures 8, 9 and 10 show the transient PINN and CFD models results of velocity magnitude, static pressure and temperature for timesteps of 0.15 s, 0.5 s and 1 s respectively. In figure 8 we see that the PINN model accurately predicts the velocity evolution when compared to the CFD results for the three time steps. The results in figure 9 for 0.15 s show that the PINN model underpredicts the static pressure distribution near the inlet (+-1 Pa vs. +-2 Pa) and in between the two cylinders (+-1.25 Pa vs. +-0.75 Pa) when compared to the CFD results. Similarly, for the 0.5 s results the PINN model underpredicts the static pressure, but for the 1 s results it is seen that the PINN model accurately captures the CFD pressure distribution.

The temperature contours at 0.15 and 0.5 s show that the PINN model again predicts a slightly thicker thermal boundary layer adjacent to all the walls, especially on the horizontal walls near the inlet (0.0 – 0.4 m). At 1 s, it is seen that the temperature distribution predicted by PINN model in the wakes of the cylinders is slightly more diffuse when compared to the CFD results. A potential cause for this could be that there are not enough colocation coordinates in these regions to capture the finer details to the temperature profiles. In general, the transient PINN model captures the temporal evolution of the scalar and vector field quantities with good accuracy.

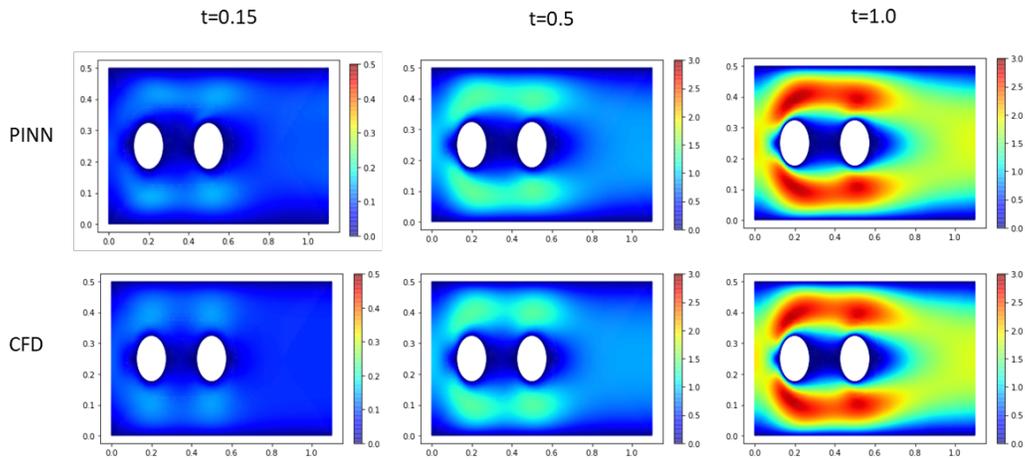

Figure 8: Comparison between transient CFD and PINN velocity predictions

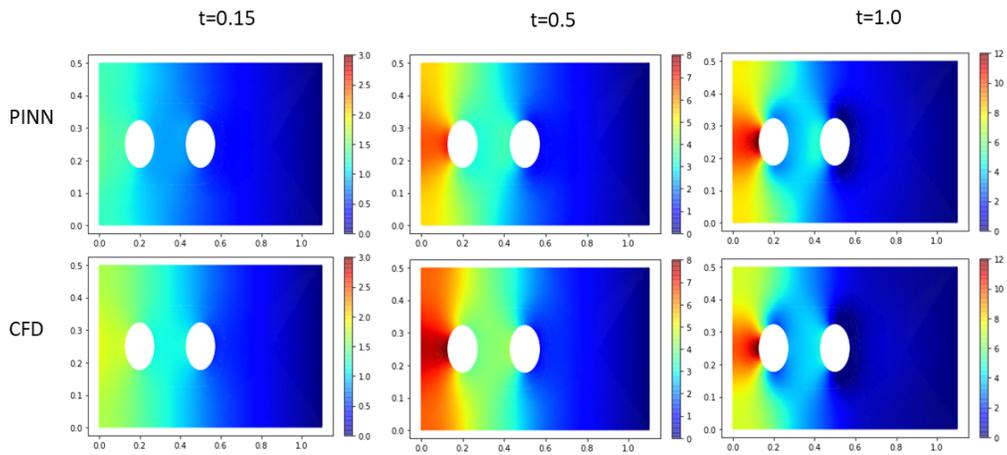

Figure 9: Comparison between transient CFD and PINN static pressure predictions

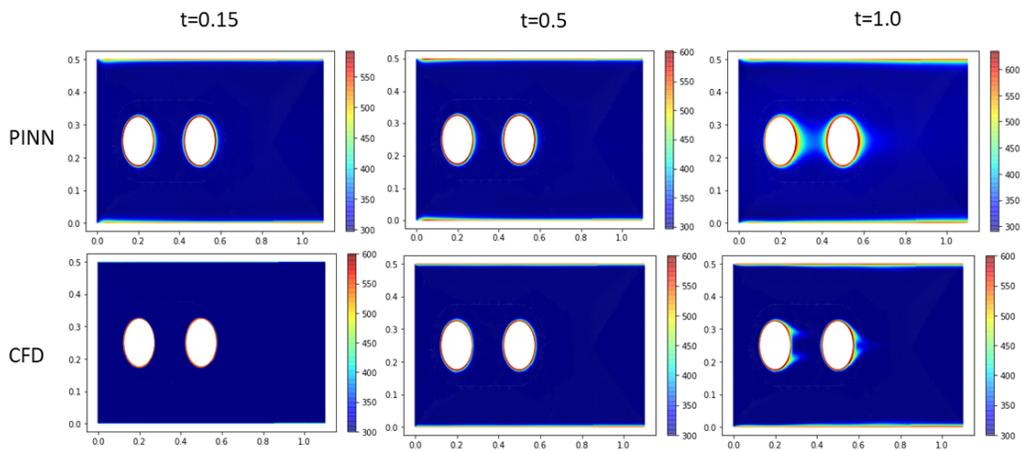

Figure 10: Comparison between transient CFD and PINN temperature predictions

Similar to the steady-state case the MAPEs for the transient results was calculated for each of the above shown time steps. The results can be seen below in table 3. The results show that the velocity, pressure and temperature errors between the predictions PINN and CFD models becomes smaller as flow time progresses, which is in agreement with what is observed in figure 8-10.

Table 3: Mean absolute percentage errors between CFD and PINN model predictions for different time steps

| Solution variable | Velocity magnitude, [%] | Static pressure, [%] | Temperature, [%] |
|---|---|---|---|
| $t = 0.15$ s | 5.1 | 18 | 4.0 |
| $t = 0.5$ s | 3.5 | 16 | 3.1 |
| $t = 1.0$ s | 2.7 | 12 | 2.9 |

## 4. Conclusions

In the present work, the mixed-variable PINN approach was applied to model steady-state and transient mass, momentum and energy conservation in a simple 2D internal flow domain with immersed heated cylinders and duct walls. The PINN solved the relevant physics PDEs in their respective non-dimensionalized form to circumvent large differences between conservation equation loss values calculated during training. The non-dimensionalized approach showed a significant improvement over simply using the dimensional form of the PDEs. For both the transient and steady-state configurations the agreement between the PINN and CFD results are good, showcasing the ability of the PINN approach to solve multiple physics-based PDEs on a single domain.

It must be stated that the PINN approach takes significantly longer to solve the PDEs compared to the traditional finite volume numerical methods utilized in commercial CFD software. For example, the steady-state CFD model solved within 5 minutes using an i7-7700K quad-core processor whereas the PINN method took approximately an hour to train using a Nvidia RTX 2080 GPU card. Once the PINN is trained it can generate the predictions almost instantaneously. Therefore, as mentioned in [4], PINNs should not be seen as a replacement for traditional numerical PDE solvers. Rather, it can be applied as a modelling methodology in areas such as development of computationally efficient sub-models which can be integrated into traditional CFD solvers and to be used in model predictive control optimization routines. A possible example of the former is to use PINNs as a fine structure chemical kinetics integrator in the EDC turbulence-chemistry interaction model to speed-up solving times of complex combustion problems [29].

Future work entails investigating the addition of variable boundary condition values to the input vectors supplied to the PINN during training to enable solving of different cases with a single network, turbulence modelling, radiation transport modelling and application of PINNs to solve 1D process models of thermofluid networks, for example heat pump systems and heat exchanger networks.

## 5. Acknowledgements

Funding: The authors would like to thank the Eskom EPPEI program for funding this project.